\documentclass[12pt,aps,prd,preprint,groupedaddress,tightenlines,nofootinbib,floatfix]{revtex4}

\usepackage[dvipdfm]{graphicx}

\begin{document}

\begin{flushright}
{\normalsize UTHEP-655}\\
{\normalsize UTCCS-P-69}\\
\end{flushright}

\title{Charmed baryons at the physical point in 2+1 flavor lattice QCD}

\author{Y.~Namekawa$^1$, S.~Aoki$^{1,2}$, K.~-I.~Ishikawa$^3$, N.~Ishizuka$^{1,2}$,
K.~Kanaya$^2$, Y.~Kuramashi$^{1,2,4}$,
M.~Okawa$^3$, Y.~Taniguchi$^{1,2}$,
A.~Ukawa$^{1,2}$, N.~Ukita$^1$ and T.~Yoshi\'e$^{1,2}$\\
(PACS-CS Collaboration)
}

\affiliation{
$^1$~Center for Computational Sciences, University of Tsukuba,
Tsukuba, Ibaraki 305-8577, Japan\\
$^2$~Graduate School of Pure and Applied Sciences, University of Tsukuba,
Tsukuba, Ibaraki 305-8571, Japan\\
$^3$~Graduate School of Science, Hiroshima University,
Higashi-Hiroshima, Hiroshima 739-8526, Japan\\
$^4$~RIKEN Advanced Institute for Computational Science,
Kobe, Hyogo 650-0047, Japan
}

\date{\today}

\begin{abstract}
We investigate the charmed baryon mass spectrum
using the relativistic heavy quark action 
on 2+1 flavor PACS-CS configurations
previously generated on
$32^3 \times 64$ lattice.  
The dynamical up-down and strange quark masses
are tuned to their physical values,
reweighted from those employed in the configuration generation.
At the physical point, the inverse lattice spacing
determined from the $\Omega$ baryon mass
gives $a^{-1}=2.194(10)$~GeV,
and thus the spatial extent becomes $L = 32 a = 2.88(1)$~fm. 
Our results for the charmed baryon masses
are consistent with experimental values, 
except for the mass of $\Xi_{cc}$,
which has been measured by only one experimental group so far
and has not been confirmed yet by others.
In addition, we report values of other doubly and triply charmed baryon masses,
which have never been measured experimentally.
\end{abstract}

\maketitle

\setcounter{equation}{0}
\section{Introduction}
\label{section:introduction}

Recently, a lot of new experimental results are
reported on charmed baryons~\cite{PDG_2012}.
BaBar and Belle give very accurate results based on their precise analysis. 
In addition, new experiments
such as J-PARC, PANDA, LHCb, and Belle II
are coming and expected to give further 
informations for charmed baryons.

Mass spectrum of singly charmed baryons has been
determined experimentally in high accuracy.
Experimental status for masses of the ground state
is evaluated as three or four-star by the particle data group.
The excited states are also investigated fairly well.

In contrast to singly charmed baryons,
experimental data for doubly and triply charmed baryons
are not well established.
A candidate for the doubly charmed baryon, $\Xi_{cc}$,
has been reported only by the SELEX Collaboration~\cite{SELEX_2002_2005},
while $\Xi_{cc}$ has not been confirmed yet
by other experimental groups such as
BaBar~\cite{BaBar_2006} and Belle~\cite{Belle_2006}.
Further experimental and theoretical confirmations are required
to establish $\Xi_{cc}$.
Other doubly charmed baryons and any triply charmed baryons
have not been observed by experiments.
In this situation, 
theoretical predictions for the doubly and triply charmed baryon masses
may give useful information for the experimental discovery of these states.

Charmed baryon spectrum has been mainly investigated
using gauge configurations generated with 2+1 flavors
dynamical staggered quarks~\cite{Na_2006,Liu_2010,Briceno_2012,Mathur_2012}.
In this case, a choice for light valence quarks requires a special care:
One may take other fermion formulations for the valence light quarks
(the mixed action) to avoid a problem due to a complicated flavor structure
of the staggered quarks.
Alternatively one may construct charmed baryon operators
with the valence naive quark
and then rewrite the correlation function of these operators in terms of
the staggered propagators~\cite{Wingate_2003}.
Both approaches violate the unitarity of the theory at finite lattice spacing,
in addition to the rooting problem of dynamical staggered quarks. 
Therefore it is necessary to check their results
using other combinations of dynamical and valence quarks
which maintain the unitarity at non-zero lattice spacing.
Another issue in the existing calculations on dynamical staggered configurations
is that their chiral extrapolations suffer from large higher order corrections.
NLO SU(2) heavy baryon chiral perturbation theory
is employed to extrapolate their data to the physical point
from 220 -- 290 MeV pion masses, 
but the result shows a bad convergence even at $m_{\pi} = 220$~MeV.
A calculation directly at the physical quark masses without chiral extrapolation
is the best way to remove this uncertainty.

There are a few investigations with other fermion formulations.
S.~D\"{u}rr {\it et al} have calculated charmed omega baryon masses
using smeared improved Wilson and Brillouin fermions
for valence strange and charm quarks~\cite{Durr_2012}.
Though their calculation is performed only at one quark mass ($m_{\pi}=280$~MeV)
and one lattice spacing ($a=0.07$~fm) on the $N_f=2$ $O(a)$-improved Wilson quark ensemble, 
they obtained a result which is consistent with the experimental value
for the singly charmed omega baryon ($\Omega_c$) mass.

The ETMC has studied charmed baryons on ensembles
generated with $N_f=2$ dynamical twisted-mass quarks,
employing the same twisted-mass fermion for degenerate up and down valence quarks
and Osterwalder-Seiler fermions for strange and charm valence quarks~\cite{ETMC_2012}.
For a doubly charmed baryon,
they have found $m_{\Xi_{cc}} = 3.513(23)(14)$ GeV.
This is the only result that is consistent with the SELEX experimental value
$m_{\Xi_{cc}}^{\rm SELEX} = 3.519(1)$~GeV,
while other lattice QCD results deviate from this value.
Reasons for this disagreement among lattice QCD results
must be understood and should be eventually resolved.
One possible source for systematic uncertainties in the ETMC calculation
is a lattice artifact caused by the heavy charm quark mass
at their lattice spacings, $a = 0.09-0.06$~fm.
Indeed their results for charmed baryon masses, especially for $m_{\Omega_{ccc}}$,
do not show clear scaling behaviors.
To reduce this uncertainty, one must employ a heavy quark action
that handles mass dependent lattice artifacts in the formulation,
such as the Fermilab action~\cite{Fermilab_action},
the relativistic heavy quark action~\cite{RHQ_action_Tsukuba,RHQ_action_Columbia},
or highly improved actions.
Chiral extrapolation of ETMC data from $m_{\pi}=260$~MeV
using the NLO heavy baryon chiral perturbation theory
is another source of systematic uncertainties,
as in the case of staggered quarks.

In Ref.~\cite{PACS_CS_2011},
the present authors have shown that the charm quark mass corrections
are under control at $a^{-1}=2.194(10)$~GeV
by adopting the relativistic heavy quark action of Ref.~\cite{RHQ_action_Tsukuba}.
It removes the leading cutoff errors of $O((m_Q a)^n)$
and the next to leading effects of $O((m_Q a)^n (a \Lambda_{QCD}))$
for arbitrary order $n$ by tuning finite number of parameters.
Employing this action for the charm quark,
we have investigated properties of mesons involving charm quarks
with the 2+1 dynamical flavor PACS-CS configurations
on $32^3 \times 64$ lattice~\cite{PACS_CS_1}
reweighted to the physical point for 
up, down and strange quark masses.
We have found our results for charmed meson masses are consistent
with experiment at a percent level,
and so are those for the decay constants with a few percent accuracy,
though our results are obtained at a single lattice spacing.

Encouraged by this result,
we have extended our investigation to the charmed baryon sector
and report the result in this paper.
One of the big advantage in our investigation over previous calculations is 
that the chiral extrapolation is no more necessary,
since our calculations are performed at the physical point.
We are free from the convergence problem
of the heavy baryon chiral perturbation theory.
We first compare our masses of singly charmed baryons
with the corresponding experimental values,
to check if our method works also for the baryon sector.
We then evaluate the doubly and triply charmed baryon spectra
as our predictions.
A part of this work has been reported in Ref.~\cite{Namekawa_2012}.

This paper is organized as follows.
Section II explains our method and simulation parameters.
Section III describes our results
for the singly charmed baryon spectrum,
and comparison with experiments.
In Sec. IV, we present our results for doubly and triply charmed baryon masses.
A conclusion is given in Sec. V.

\setcounter{equation}{0}
\section{Set up}
\label{section:setup}

Our investigation is based on a set of $2+1$ flavor
dynamical lattice QCD configurations
generated by the PACS-CS Collaboration~\cite{PACS_CS_1}
on a $32^3\times 64$ lattice 
using the nonperturbatively $O(a)$-improved Wilson quark action 
with $c_{\rm SW}^{\rm NP}=1.715$~\cite{Csw_NP}
and the Iwasaki gauge action~\cite{RG}
at $\beta=1.90$.
The aggregate of 2000 MD time units were generated at the hopping parameter
given by $(\kappa_{ud}^0,\kappa_{s}^0)=(0.13778500, 0.13660000)$,
and 80 configurations separated by 25 MD time units were selected for our calculations.
We then reweight those configurations to the physical point
given by $(\kappa_{ud},\kappa_{s})=(0.13779625, 0.13663375)$.
The reweighting shifts the masses of $\pi$ and $K$ mesons 
from $m_\pi=152(6)$~MeV and $m_K=509(2)$~MeV
  to $m_\pi=135(6)$~MeV and $m_K=498(2)$~MeV,
with the cutoff at the physical point estimated
to be $a^{-1}=2.194(10)$~GeV from the $\Omega$ baryon mass.
Our parameters and statistics
at the physical point are given in Table~\ref{table:statistics}.

\begin{table}[t]
\begin{center}
\begin{tabular}{ccccc}
\hline
 $\beta$           &
 $\kappa_{\rm ud}$ & $\kappa_{\rm s}$ &
 \# conf           & MD time
\\ \hline
 1.90              &
 0.13779625        & 0.13663375 &
 80                & 2000
\\ \hline
\end{tabular}
\caption{Simulation parameters.
         MD time is defined as the number of trajectories
         multiplied by the trajectory length.
}
\label{table:statistics}
\end{center}
\end{table}

The relativistic heavy quark action~\cite{RHQ_action_Tsukuba}
is designed to reduce cutoff errors of $O((m_Q a)^n)$
with arbitrary order $n$ to $O(f(m_Q a)(a \Lambda_{QCD})^2)$,
once all of the parameters in the action
are determined nonperturbatively,
where $f(m_Q a)$ is an analytic function
around the massless point $m_Q a = 0$.
The action is given by
\begin{eqnarray}
 S_Q
 &=& \sum_{x,y}\overline{Q}_x D_{x,y} Q_y,\\
 D_{x,y}
 &=& \delta_{xy}
     - \kappa_{Q}
       \sum_i \left[  (r_s - \nu \gamma_i)U_{x,i} \delta_{x+\hat{i},y}
                     +(r_s + \nu \gamma_i)U_{x,i}^{\dag} \delta_{x,y+\hat{i}}
              \right]
     \nonumber \\
 &&  - \kappa_{Q}
              \left[  (1   -     \gamma_4)U_{x,4} \delta_{x+\hat{4},y}
                     +(1   +     \gamma_4)U_{x,4}^{\dag} \delta_{x,y+\hat{4}}
              \right]
     \nonumber \\
 &&  - \kappa_{Q}
              \left[   c_B \sum_{i,j} F_{ij}(x) \sigma_{ij}
                     + c_E \sum_i     F_{i4}(x) \sigma_{i4}
              \right] \delta_{xy}.
\end{eqnarray}
The parameters $r_s, c_B, c_E$ and $\nu$
have been tuned in Ref.~\cite{PACS_CS_2011}.
It should be noticed that,
while the perturbative estimate are used for $r_s, c_B, c_E$~\cite{Aoki_2003},
the parameter $\nu$ is determined non-perturbatively
to reproduce the relativistic dispersion relation for
the spin-averaged $1S$ state of the charmonium.
The heavy quark hopping parameter $\kappa_Q$
is set to reproduce the experimental value of the mass
for the spin-averaged $1S$ state.
Our parameters for the relativistic heavy quark action
are summarized in Table~\ref{table:input_parameters_for_RHQ}.

\begin{table}[t]
\begin{center}
\begin{tabular}{cccccc}
\hline
 $\kappa_{\rm charm}$  & $\nu$     & $r_s$     & $c_B$     & $c_E$
\\ \hline
 0.10959947            & 1.1450511 & 1.1881607 & 1.9849139 & 1.7819512
\\ \hline
\end{tabular}
\caption{Parameters for the relativistic heavy quark action.
}
\label{table:input_parameters_for_RHQ}
\end{center}
\end{table}

We employ the relativistic operators for charmed baryon,
simply because the relativistic heavy quark action is employed in our calculation.
Charmed baryons can be classified under
$4 \times 4 \times 4 = 20 + 20_1 + 20_2 + \bar{4}$.
In addition to $J = 3/2$ decuplet-type 20-plet,
there are $J = 1/2$ octet-type 20-plet and $\bar{4}$-plet.

$J = 1/2$ octet-type baryon operators are given by
\begin{eqnarray}
 O_{\alpha}^{fgh}(x)
 &=& \epsilon^{abc}
     ( (q_f^a(x))^T C \gamma_5 q_g^b(x) ) q_{h \alpha}^c(x),\\
 &&  C = \gamma_4 \gamma_2,
\end{eqnarray}
where $f,g,h$ are quark flavors and $a,b,c$ are quark colors.
$\alpha = 1,2$ labels the $z$-component of the spin.
The $\Sigma$-type and $\Lambda$-type are distinguished as
\begin{eqnarray}
 \Sigma{\rm  -type}  &:& - \frac{ O^{[fh]g} + O^{[gh]f} }
                                { \sqrt{2} },
 \\
 \Lambda{\rm -type}  &:&   \frac{ O^{[fh]g} - O^{[gh]f} - 2 O^{[fg]h} }
                                { \sqrt{6} },
\end{eqnarray}
where $O^{[fg]h} = O^{fgh} - O^{gfh}$.

The decuplet-type $J = 3/2$ baryon operators are
expressed as,
\begin{eqnarray}
 D_{3/2}^{fgh}(x)
 &=& \epsilon^{abc}
     ( (q_f^a(x))^T C \Gamma_+ q_g^b(x) ) q_{h 1}^c(x),\\
 D_{1/2}^{fgh}(x)
 &=& \epsilon^{abc}
     [ ( (q_f^a(x))^T C \Gamma_0 q_g^b(x) ) q_{h 1}^c(x)
     \nonumber \\
 &&   -( (q_f^a(x))^T C \Gamma_+ q_g^b(x) ) q_{h 2}^c(x)
     ] / 3,\\
 D_{-1/2}^{fgh}(x)
 &=& \epsilon^{abc}
     [ ( (q_f^a(x))^T C \Gamma_0 q_g^b(x) ) q_{h 2}^c(x)
     \nonumber \\
 &&   -( (q_f^a(x))^T C \Gamma_- q_g^b(x) ) q_{h 1}^c(x)
     ] / 3,\\
 D_{-3/2}^{fgh}(x)
 &=& \epsilon^{abc}
     ( (q_f^a(x))^T C \Gamma_- q_g^b(x) ) q_{h 2}^c(x),\\
 && \Gamma_{\pm} = (\gamma_1 \mp i \gamma_2)/2, \Gamma_0 = \gamma_3.
\end{eqnarray}

\begin{figure}[t]
\begin{center}
 \includegraphics[width=75mm]{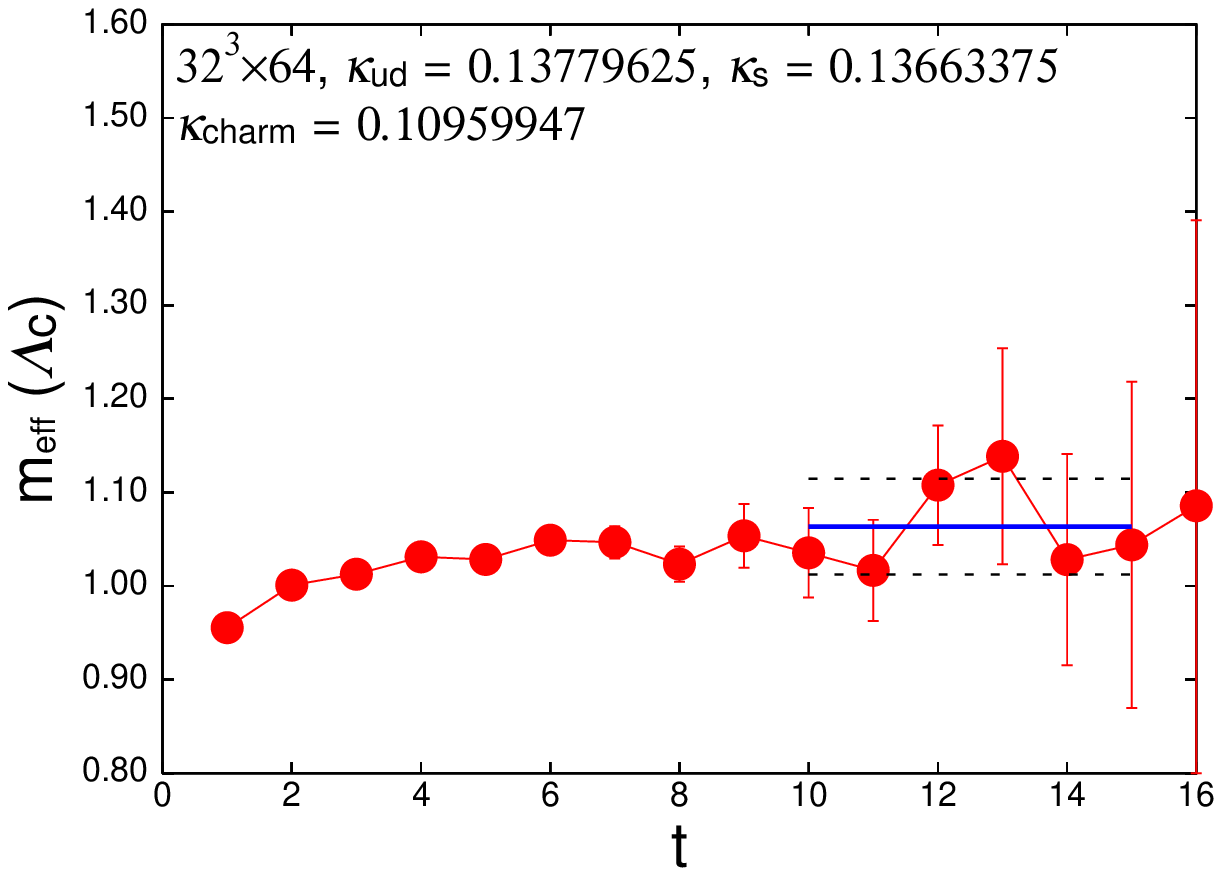}
 \includegraphics[width=75mm]{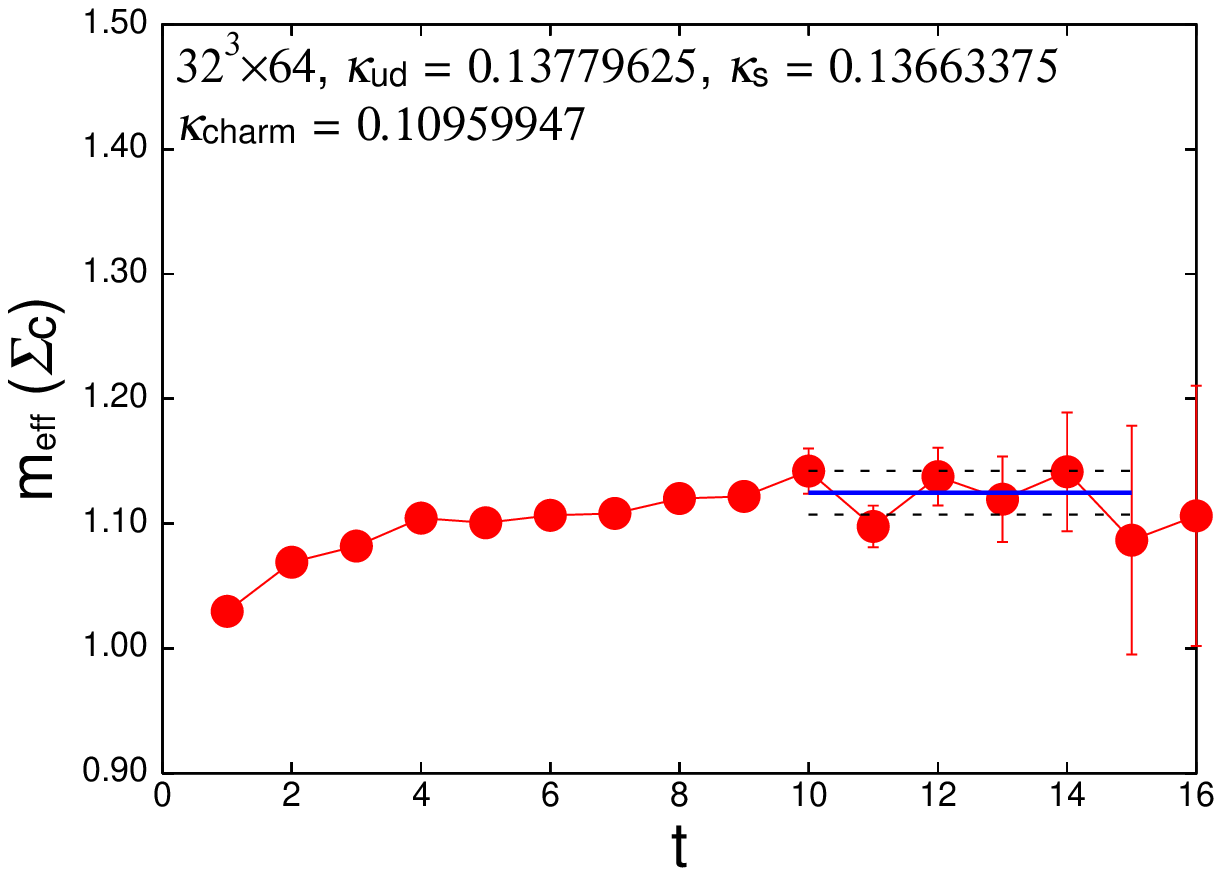}
 \includegraphics[width=75mm]{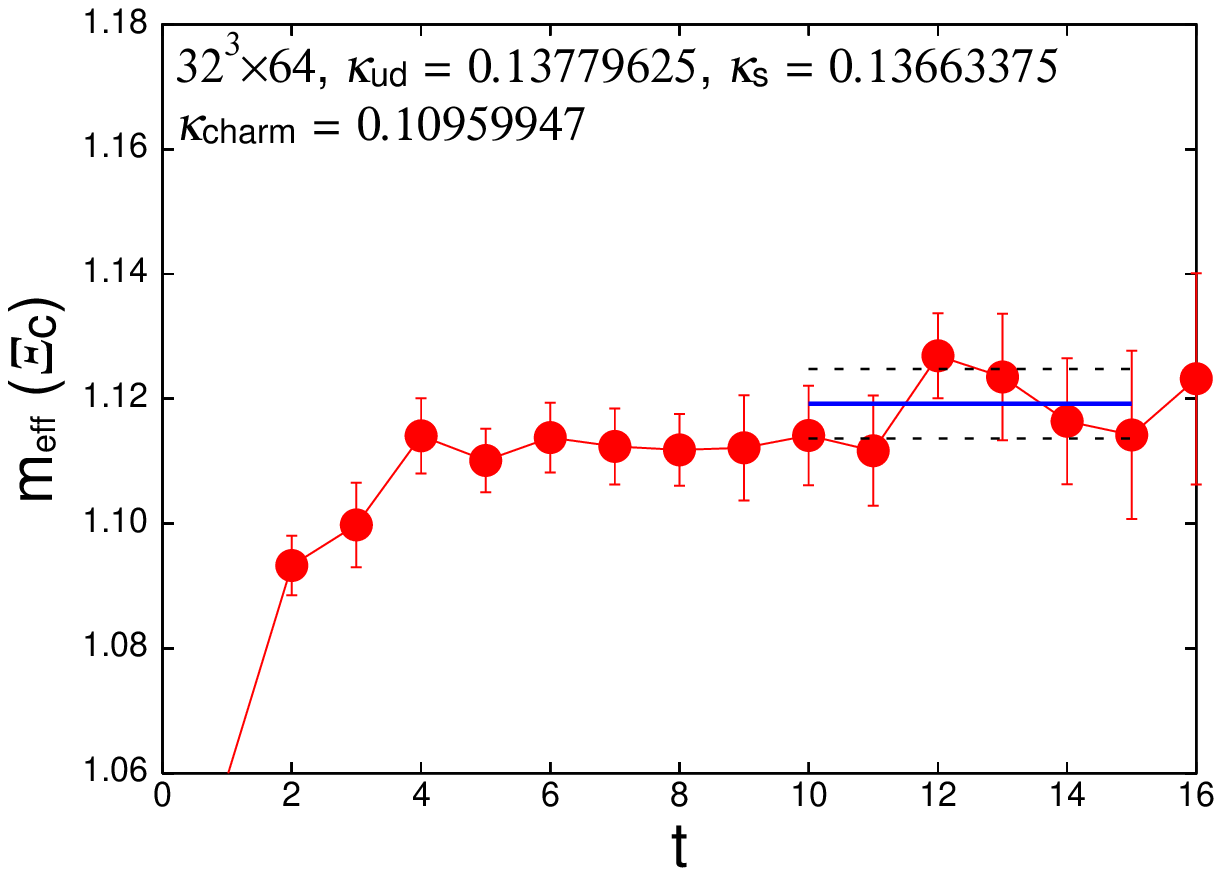}
 \includegraphics[width=75mm]{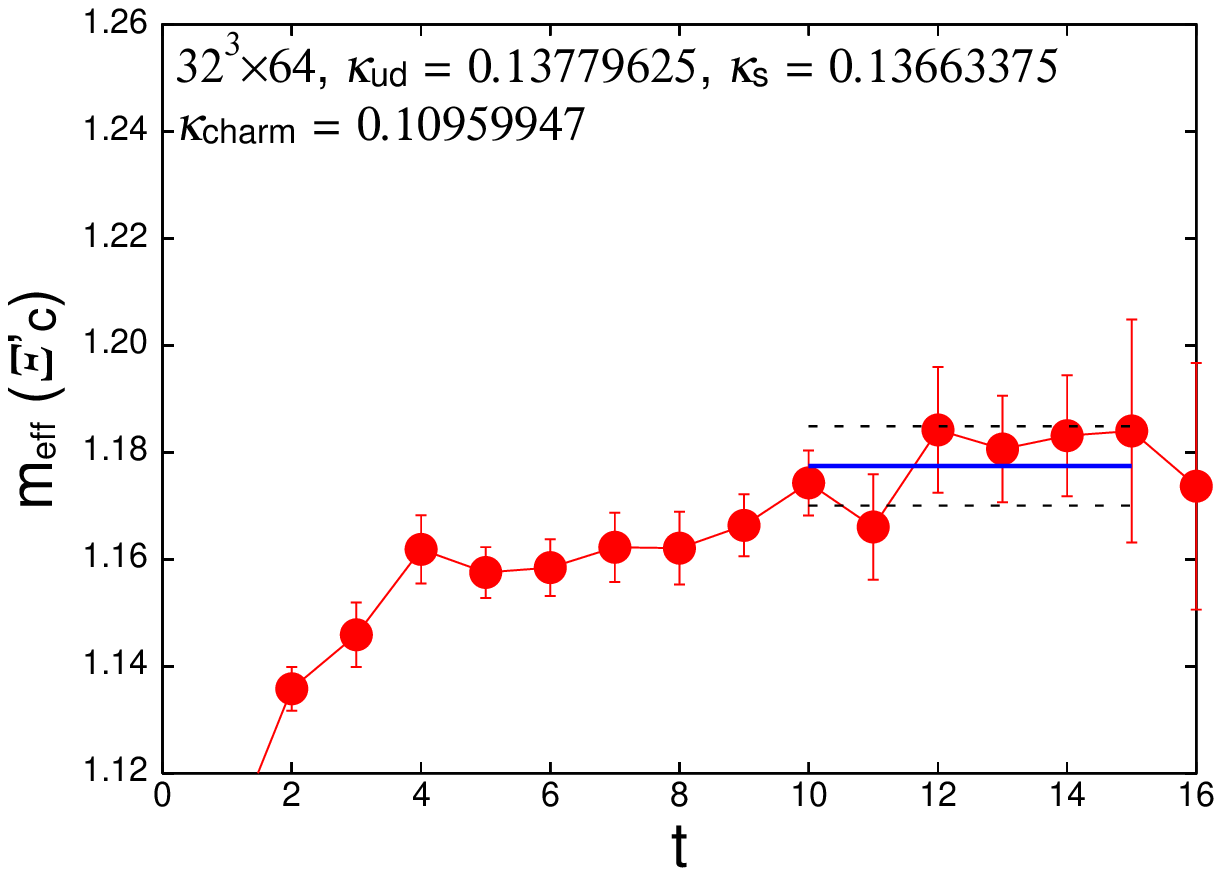}
 \includegraphics[width=75mm]{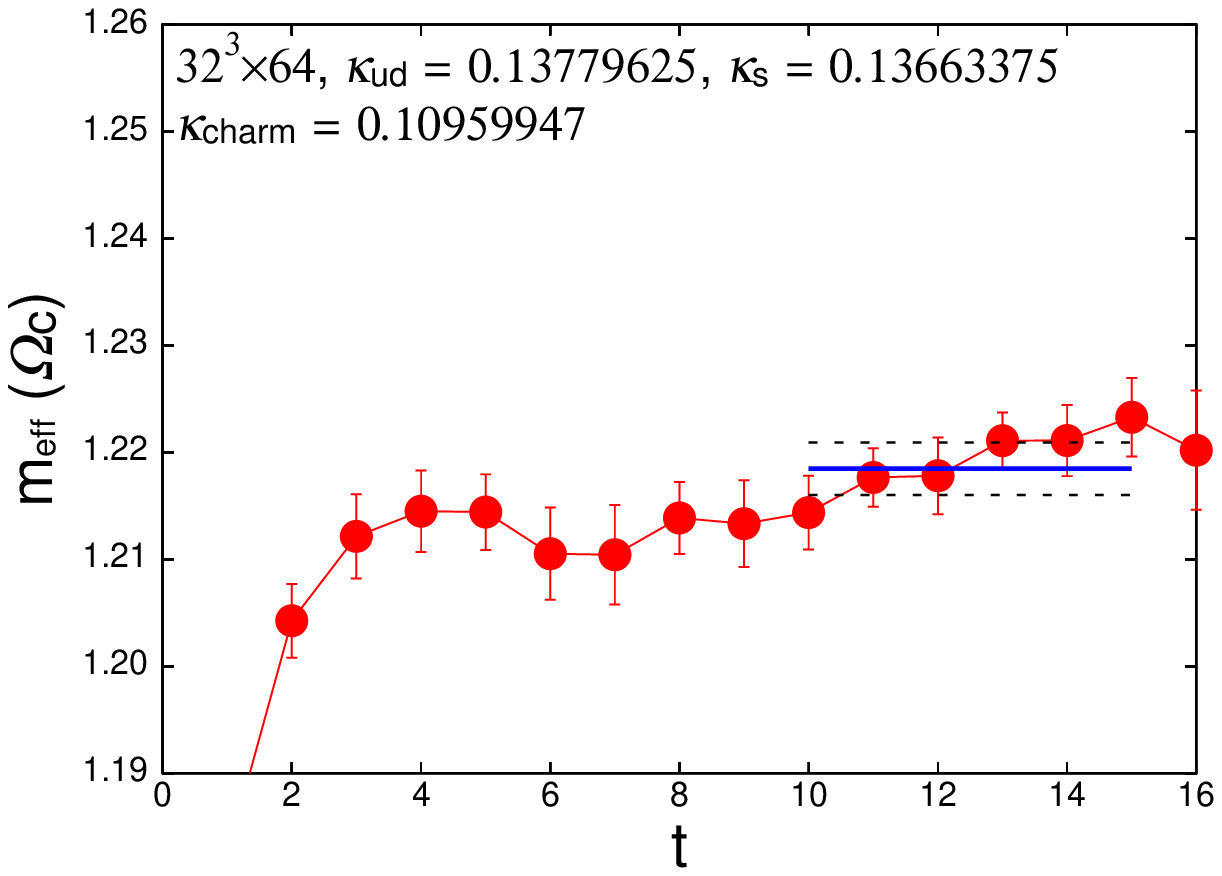}
 \caption{
  Effective masses of $J = 1/2$ singly charmed baryons.
 }
 \label{figure:m_eff_singly_1}
\end{center}
\end{figure}

\begin{figure}[t]
\begin{center}
 \includegraphics[width=75mm]{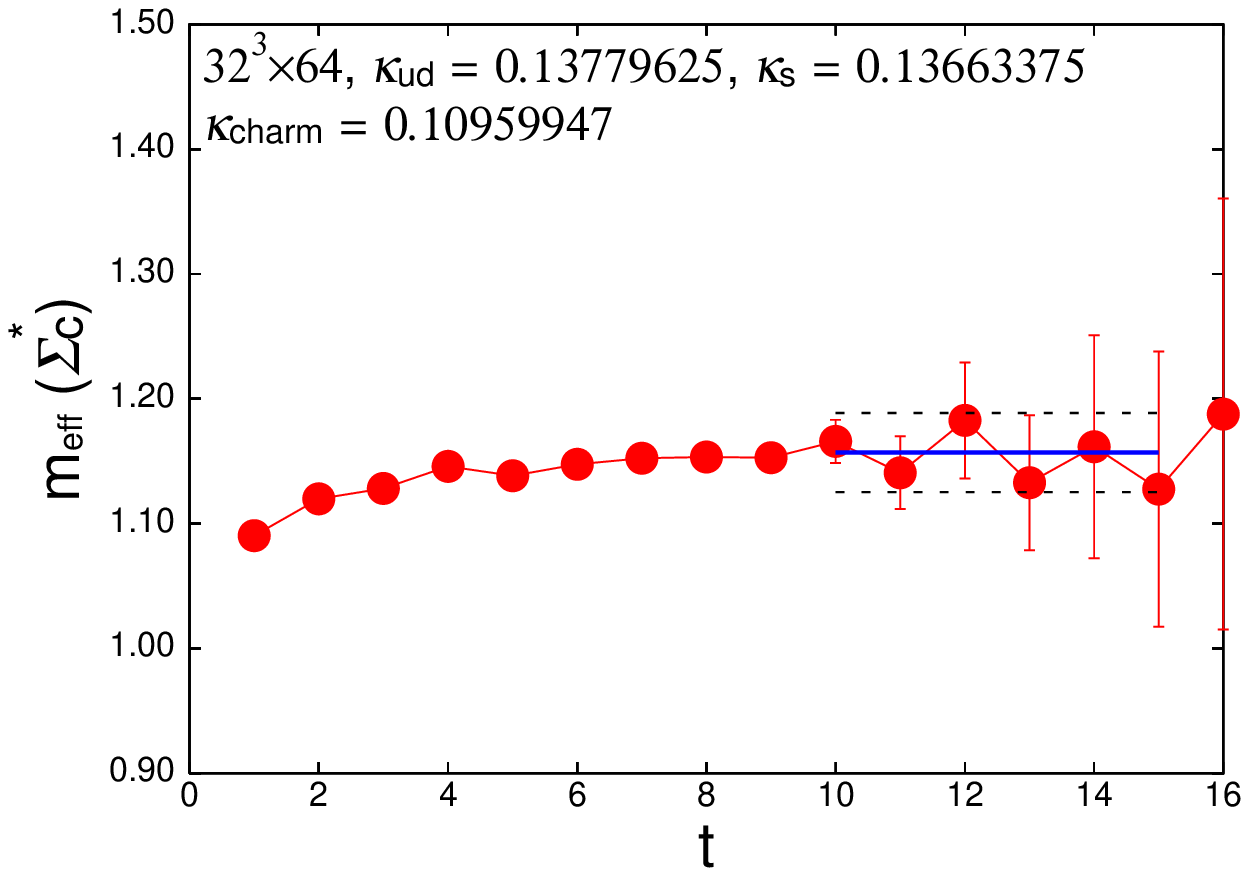}
 \includegraphics[width=75mm]{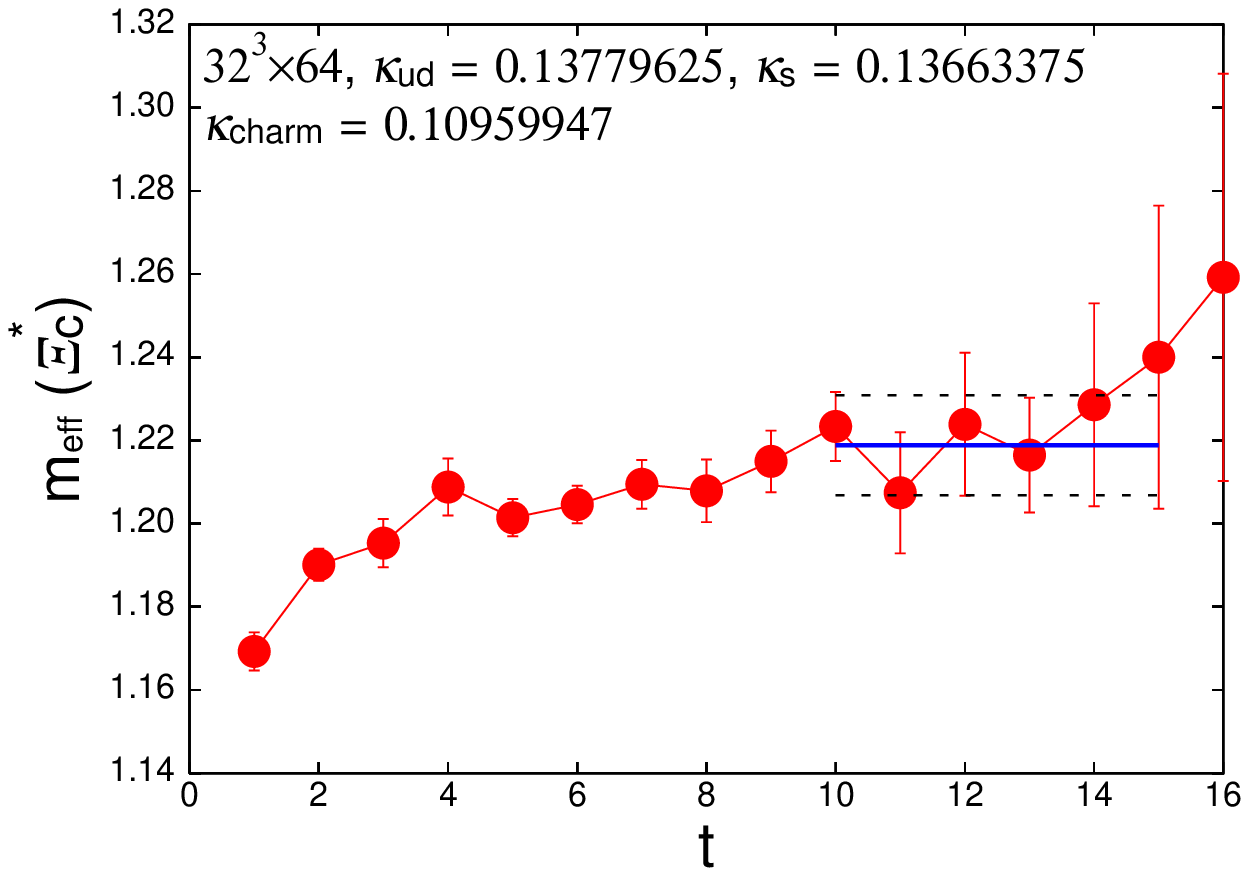}
 \includegraphics[width=75mm]{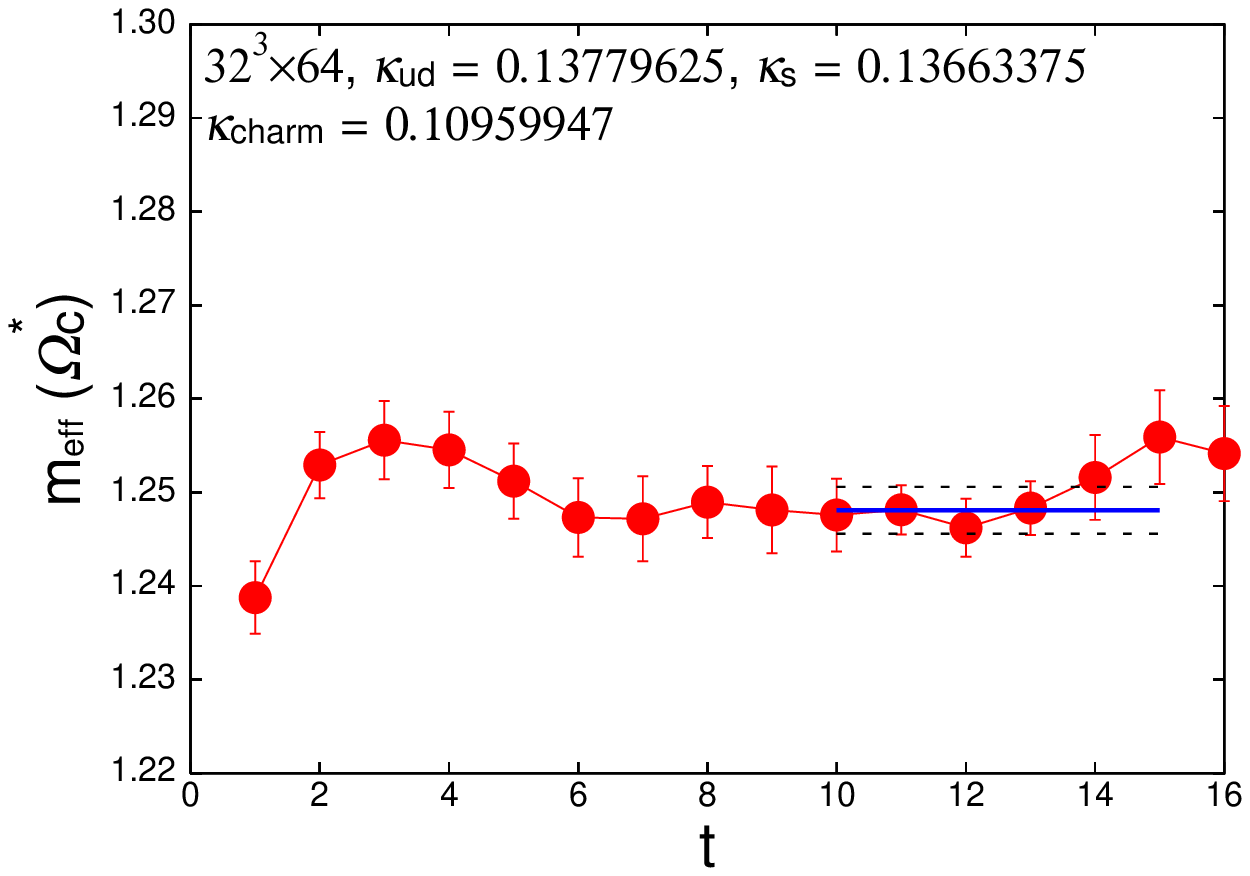}
 \caption{
  Effective masses of $J = 3/2$ singly charmed baryons.
 }
 \label{figure:m_eff_singly_2}
\end{center}
\end{figure}

\begin{figure}[t]
\begin{center}
 \includegraphics[width=75mm]{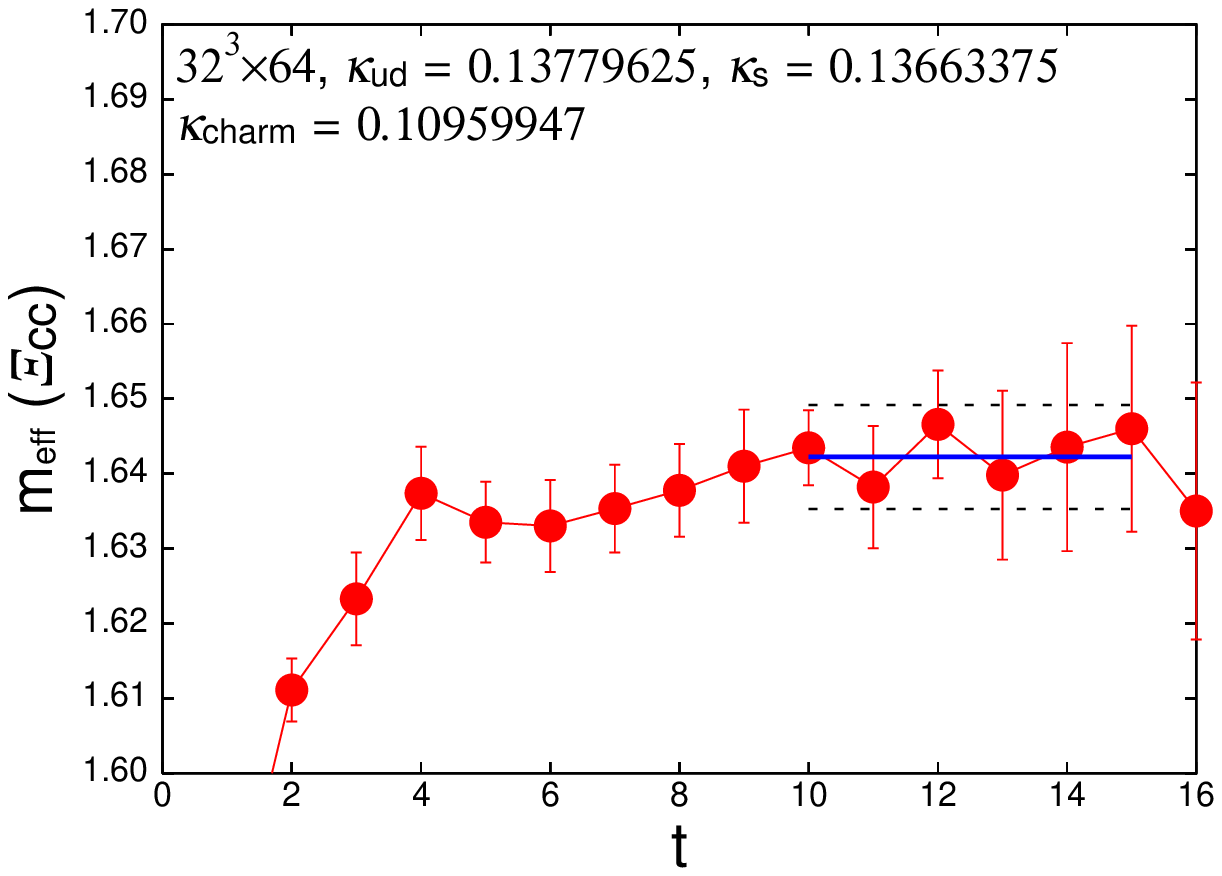}
 \includegraphics[width=75mm]{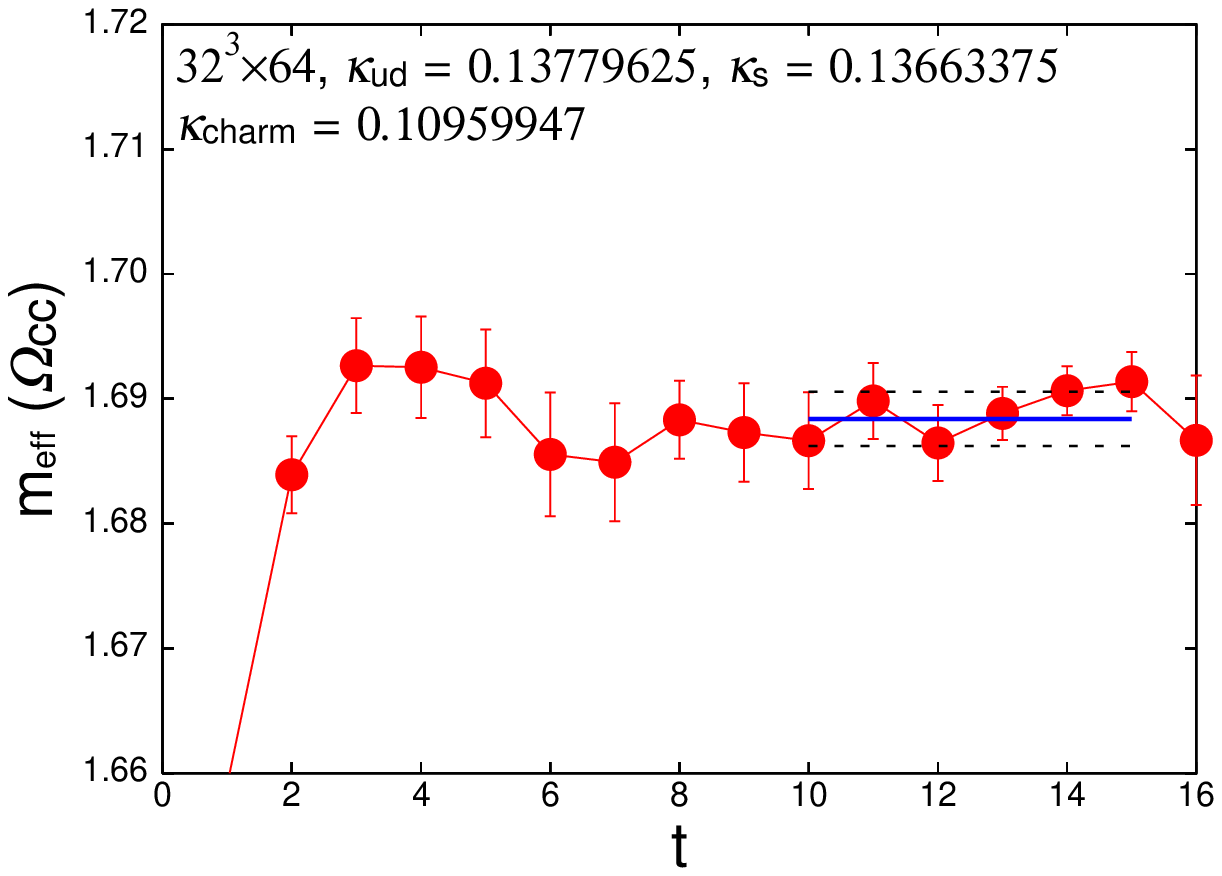}
 \includegraphics[width=75mm]{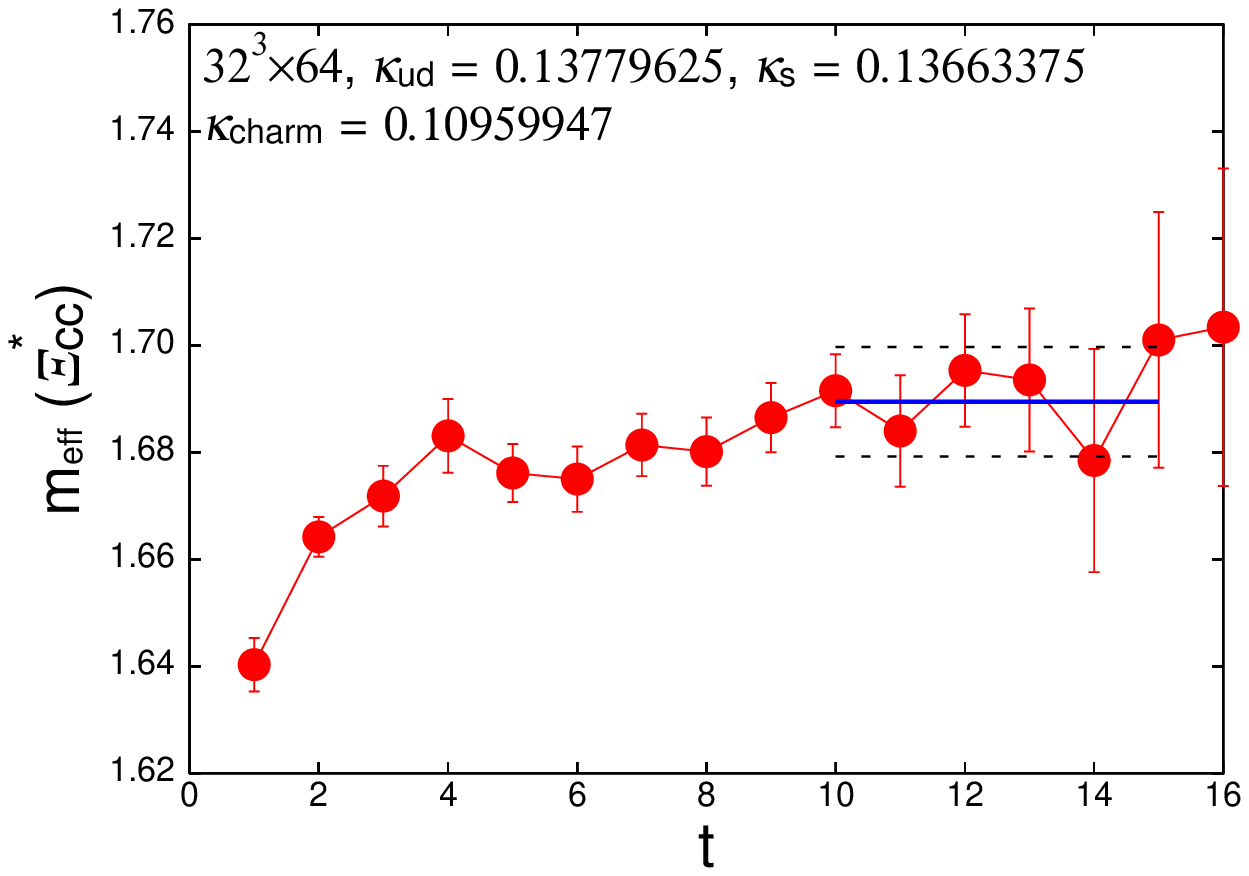}
 \includegraphics[width=75mm]{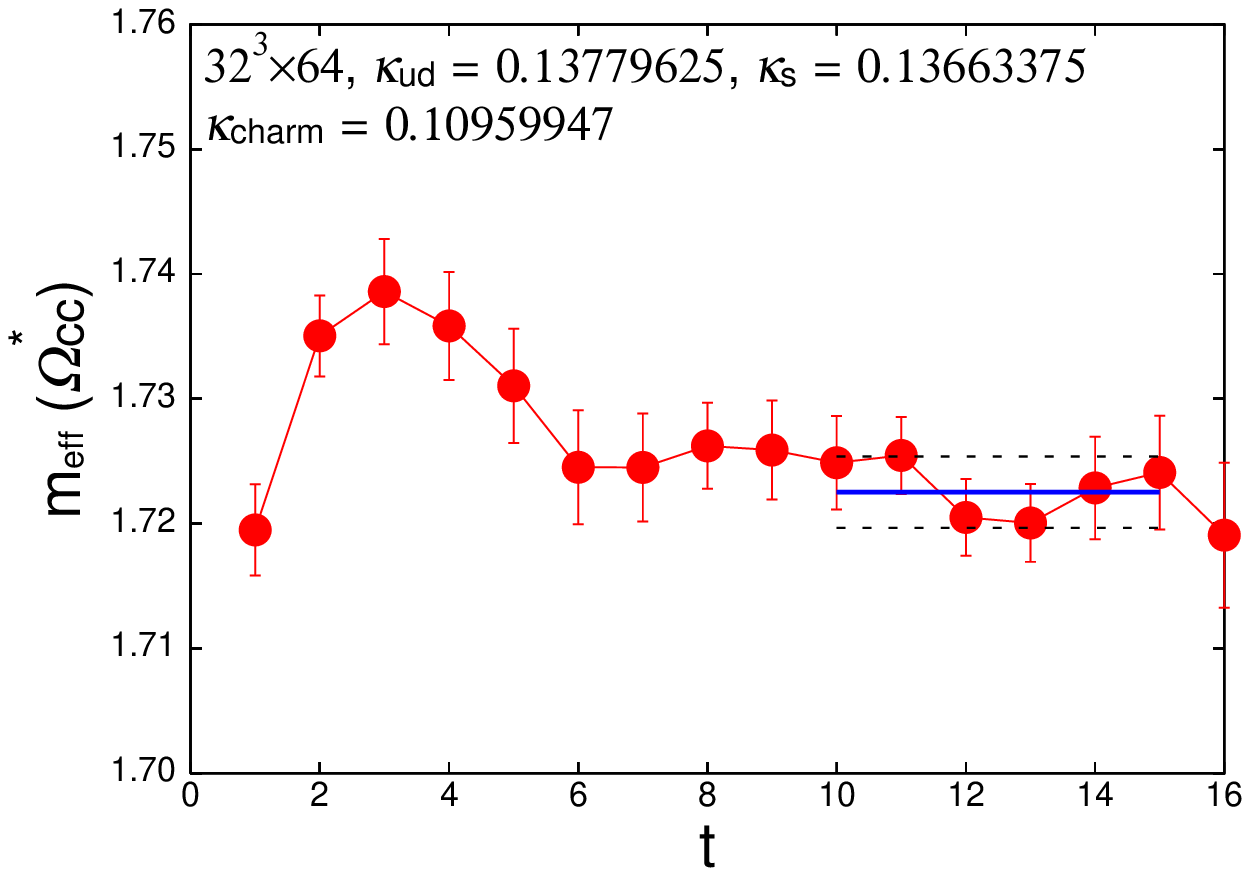}
 \includegraphics[width=75mm]{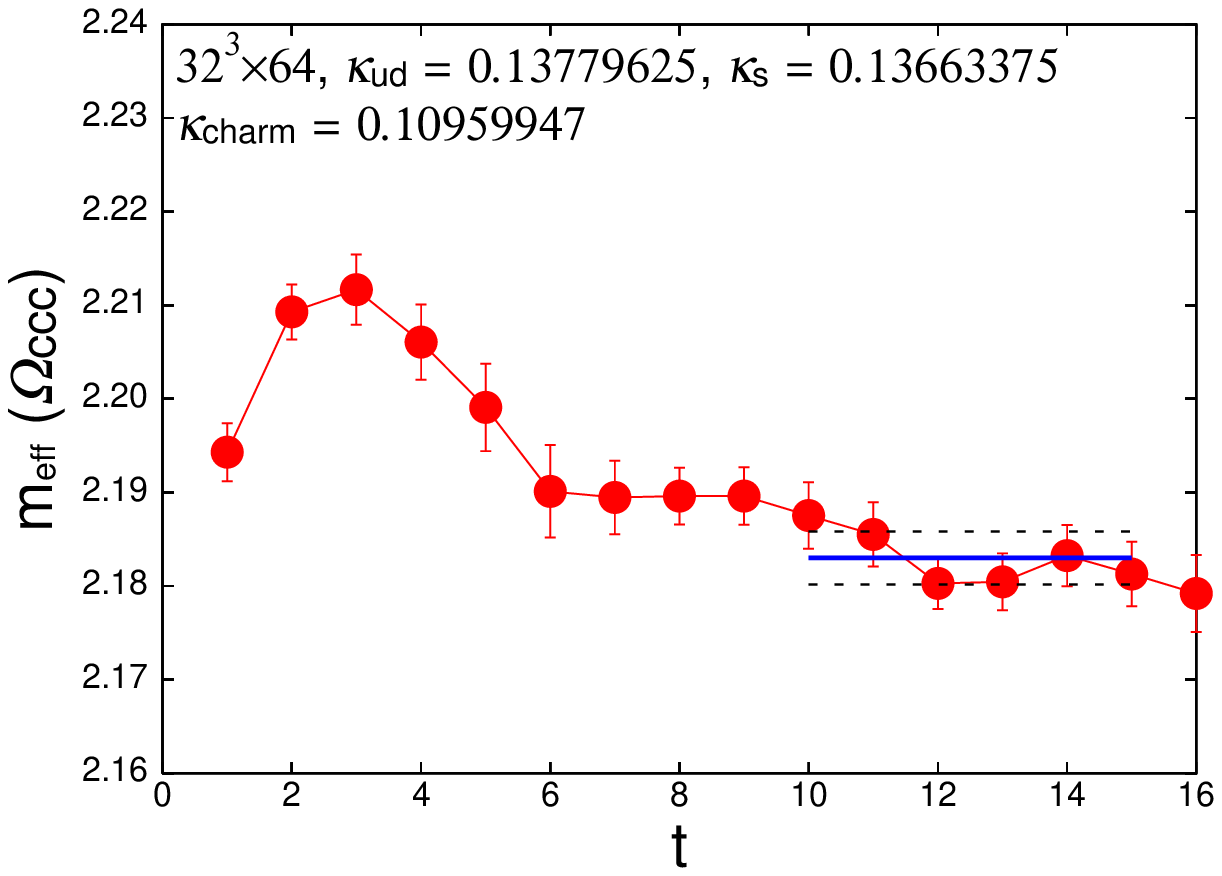}
 \caption{
  Effective masses of doubly and triply charmed baryons.
 }
 \label{figure:m_eff_doubly_and_triply}
\end{center}
\end{figure}

Two-point functions for these charmed baryon operators are calculated
with exponentially smeared sources and a local sink.
The smearing function is given by $\Psi(r) = A \exp(-B r)$
at $r \not = 0$ and $\Psi(0)=1$.
We set
$A = 1.2$, $B = 0.07$ for the $ud$ quark,
$A = 1.2$, $B = 0.18$ for the strange quark, and 
$A = 1.2$, $B = 0.55$ for the charm quark.
The number of source points is 8 per configuration
and polarization states are averaged over
to reduce statistical fluctuations.
Statistical errors are analyzed by the jackknife method
with a bin size of 100 MD time units (4 configurations),
as in the light quark sector~\cite{PACS_CS_1}.
We extract charmed baryon masses by fitting two-point functions
with single exponential forms.
Figure~\ref{figure:m_eff_singly_1}--\ref{figure:m_eff_doubly_and_triply}
show our effective masses. 
We take the fitting interval for charmed baryons
as far as possible from the origin, $[t_{min},t_{max}] = [10,15]$,
to avoid possible contaminations from excited states.
Our results are compiled in Table~\ref{table:mass_1} and \ref{table:mass_2}.

\begin{table}[t]
\begin{center}
\begin{tabular}{llccc}
\hline
                          & $J^P$           & $(I,S,C)$     &  Lattice      &  Experiment
\\ \hline
%
%
 $m_{\Lambda_c}$[GeV]     & $\frac{1}{2}^+$ & $(0  , 0,1)$  &  2.333(113)   &  2.286(0)
\\ \hline
 $m_{\Sigma_c}$[GeV]      & $\frac{1}{2}^+$ & $(1  , 0,1)$  &  2.467(40)    &  2.454(0)
\\ \hline
 $m_{\Xi_c}$[GeV]         & $\frac{1}{2}^+$ & $(1/2,-1,1)$  &  2.455(16)    &  2.471(1)
\\ \hline
 $m_{\Xi'_c}$[GeV]        & $\frac{1}{2}^+$ & $(1/2,-1,1)$  &  2.583(20)    &  2.578(3)
\\ \hline
 $m_{\Omega_c}$[GeV]      & $\frac{1}{2}^+$ & $(0  ,-2,1)$  &  2.673(13)    &  2.695(2)
\\ \hline
%
%
 $m_{\Sigma^*_c}$[GeV]    & $\frac{3}{2}^+$ & $(1  , 0,1)$  &  2.538(71)    &  2.519(1)
\\ \hline
 $m_{\Xi^*_c}$[GeV]       & $\frac{3}{2}^+$ & $(1/2,-1,1)$  &  2.674(29)    &  2.646(1)
\\ \hline
 $m_{\Omega^*_c}$[GeV]    & $\frac{3}{2}^+$ & $(0  ,-2,1)$  &  2.738(13)    &  2.766(2)
\\ \hline
\end{tabular}
\caption{
 Our results for singly charmed baryon masses.
 Experimental values from PDG~\cite{PDG_2012}
 are also listed.
}
\label{table:mass_1}
\end{center}
\end{table}

\begin{table}[t]
\begin{center}
\begin{tabular}{llccc}
\hline
                          & $J^P$           & $(I,S,C)$    &  Lattice    &  Experiment
\\ \hline
%
%
 $m_{\Xi_{cc}}$[GeV]      & $\frac{1}{2}^+$ & $(1/2, 0,2)$ &  3.603(22)  & (3.519(1))
\\ \hline
 $m_{\Omega_{cc}}$[GeV]   & $\frac{1}{2}^+$ & $(0  ,-1,2)$ &  3.704(17)  &  --
\\ \hline
%
%
 $m_{\Xi^*_{cc}}$[GeV]    & $\frac{3}{2}^+$ & $(1/2, 0,2)$ &  3.706(28)  &  --
\\ \hline
 $m_{\Omega^*_{cc}}$[GeV] & $\frac{3}{2}^+$ & $(0  ,-1,2)$ &  3.779(18)  &  --
\\ \hline
%
%
 $m_{\Omega_{ccc}}$[GeV]  & $\frac{3}{2}^+$ & $(0  , 0,3)$ &  4.789(22)  &  --
\\ \hline
\end{tabular}
\caption{
 Our results for doubly and triply charmed baryon masses.
 An experimental value from SELEX~\cite{SELEX_2002_2005}
 is also listed.
}
\label{table:mass_2}
\end{center}
\end{table}

\setcounter{equation}{0}
\section{Singly charmed baryon spectrum}
\label{section:singly_charmed_baryon}

\begin{figure}[t]
\begin{center}
 \includegraphics[width=75mm]{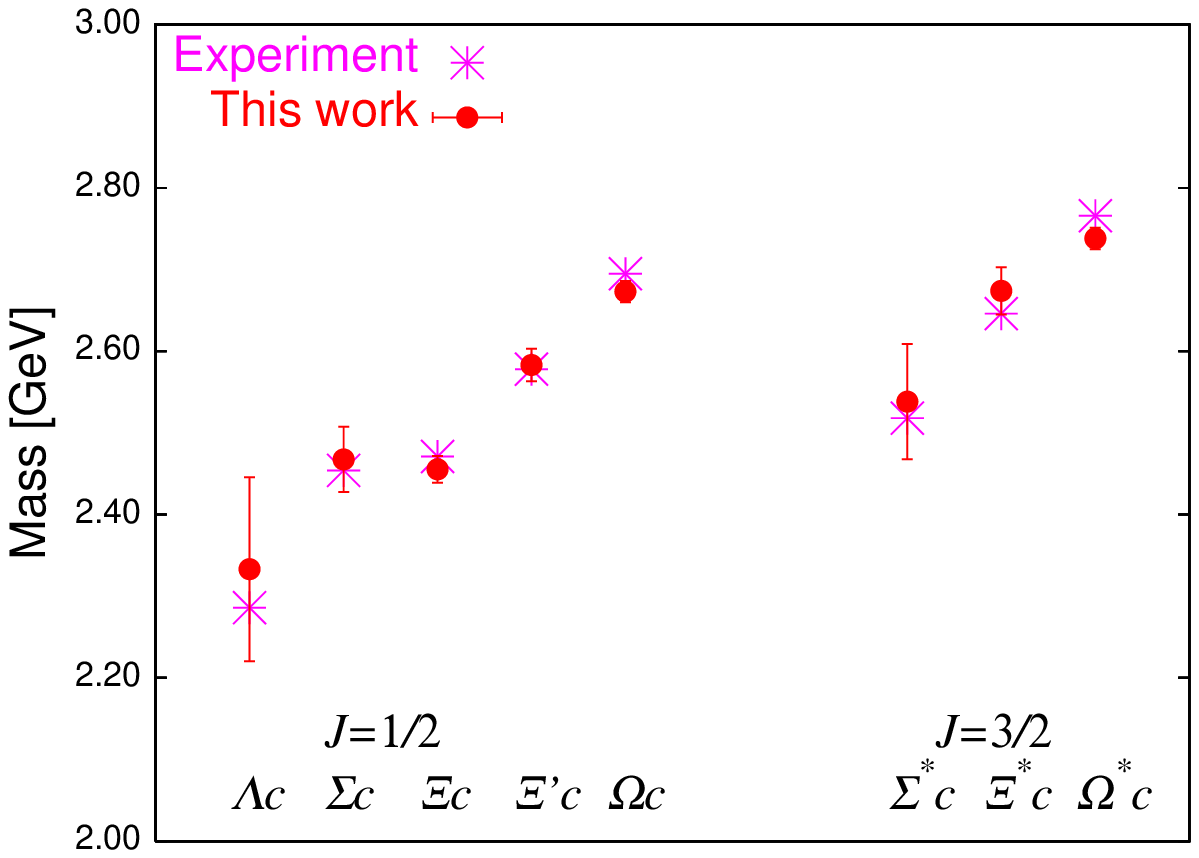}
 \includegraphics[width=75mm]{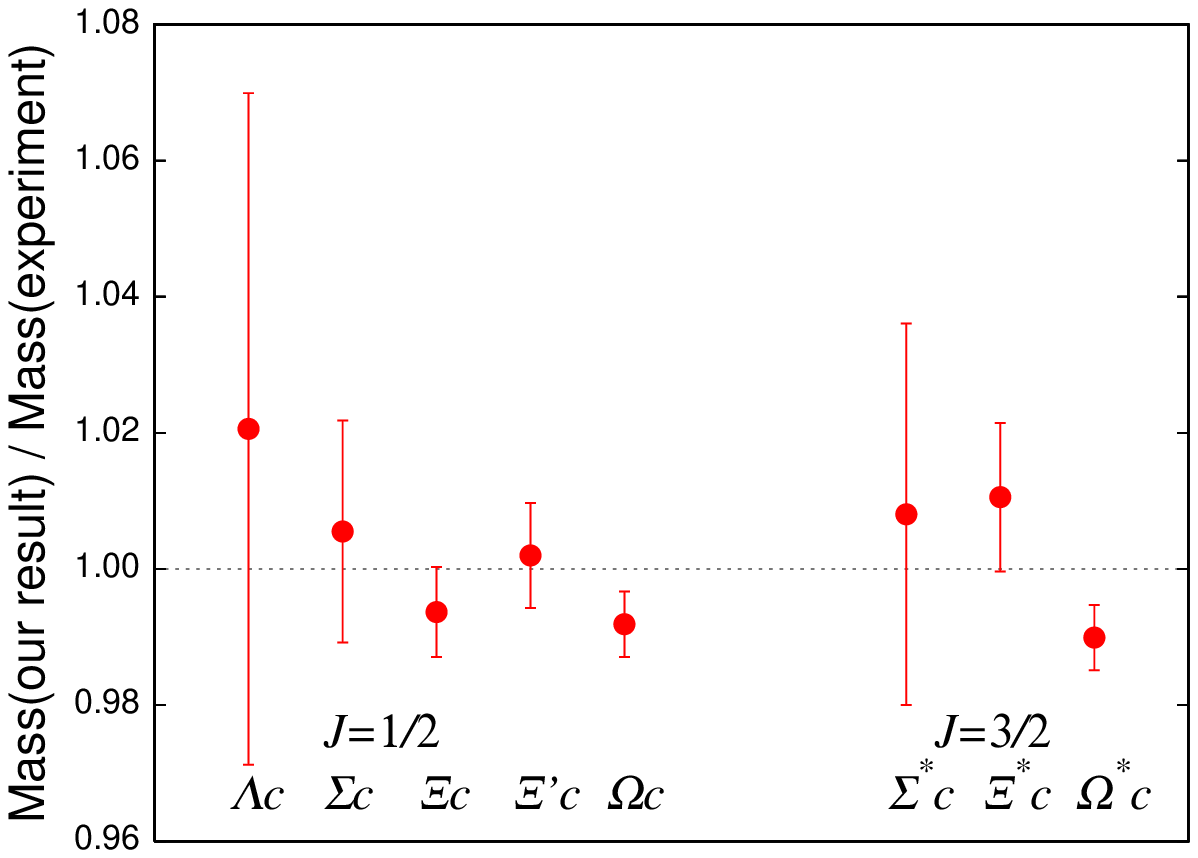}
 \caption{
  Our results for the singly charmed baryon spectrum (left panel),
  and those normalized by the experimental values (right panel).
 }
 \label{figure:mass_singly_charmed_experiment}
\end{center}
\end{figure}

Our results for the singly charmed baryon spectrum
at the physical point are summarized in
Fig.~\ref{figure:mass_singly_charmed_experiment}.
All our values for the charmed baryon masses are predictions from lattice QCD,
since the physical charm quark mass has already been fixed
by the mass for the spin-averaged $1S$ charmonium state and
no other experimental inputs for charmed baryon masses are required.

\begin{figure}[t]
\begin{center}
 \includegraphics[width=75mm]{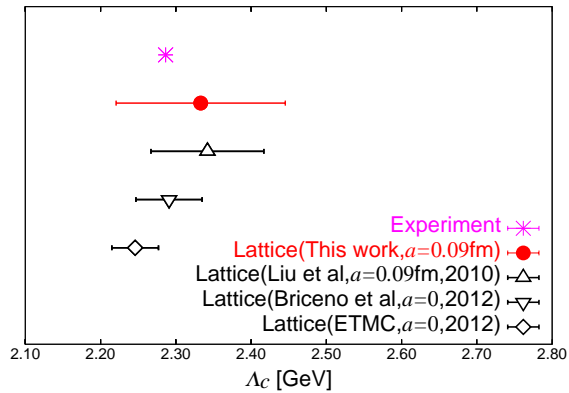}
 \caption{
 Comparison of $\Lambda_c$ mass.
 }
 \label{figure:mass_Lambda_c_lattice}
\end{center}
\end{figure}

As can be seen from the figure,
our prediction for the singly charmed baryon spectrum
in lattice QCD at a single lattice spacing 
is in reasonable agreement with the experimental one. 
In Fig.~\ref{figure:mass_Lambda_c_lattice},
we also compare our value for $\Lambda_c$
with other results obtained in recent lattice QCD simulations
using the dynamical staggered quarks~\cite{Na_2006,Liu_2010,Briceno_2012},
and the twisted mass quarks~\cite{ETMC_2012}.
All results are consistent with each other,
though the statistical error is larger for our result
due to the conservative choice of our fitting interval.

\begin{figure}[t]
\begin{center}
 \includegraphics[width=75mm]{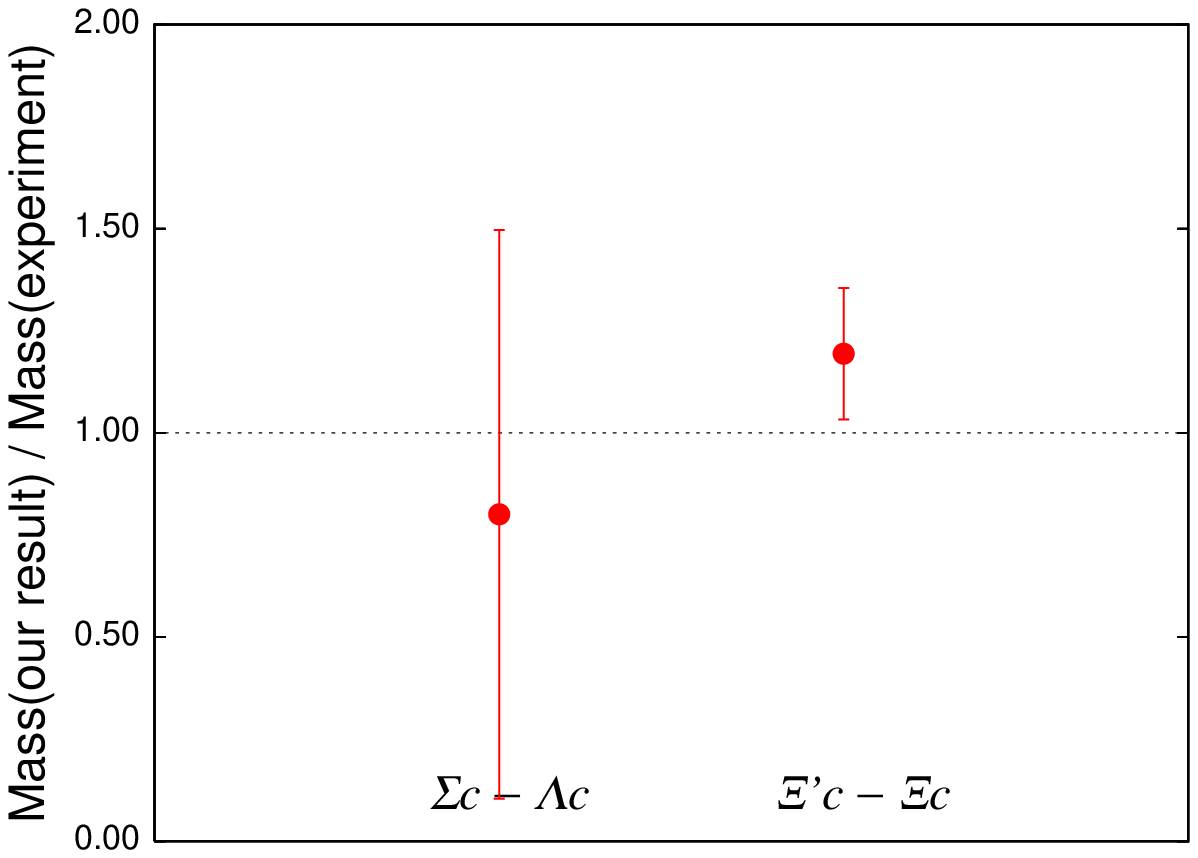}
 \includegraphics[width=75mm]{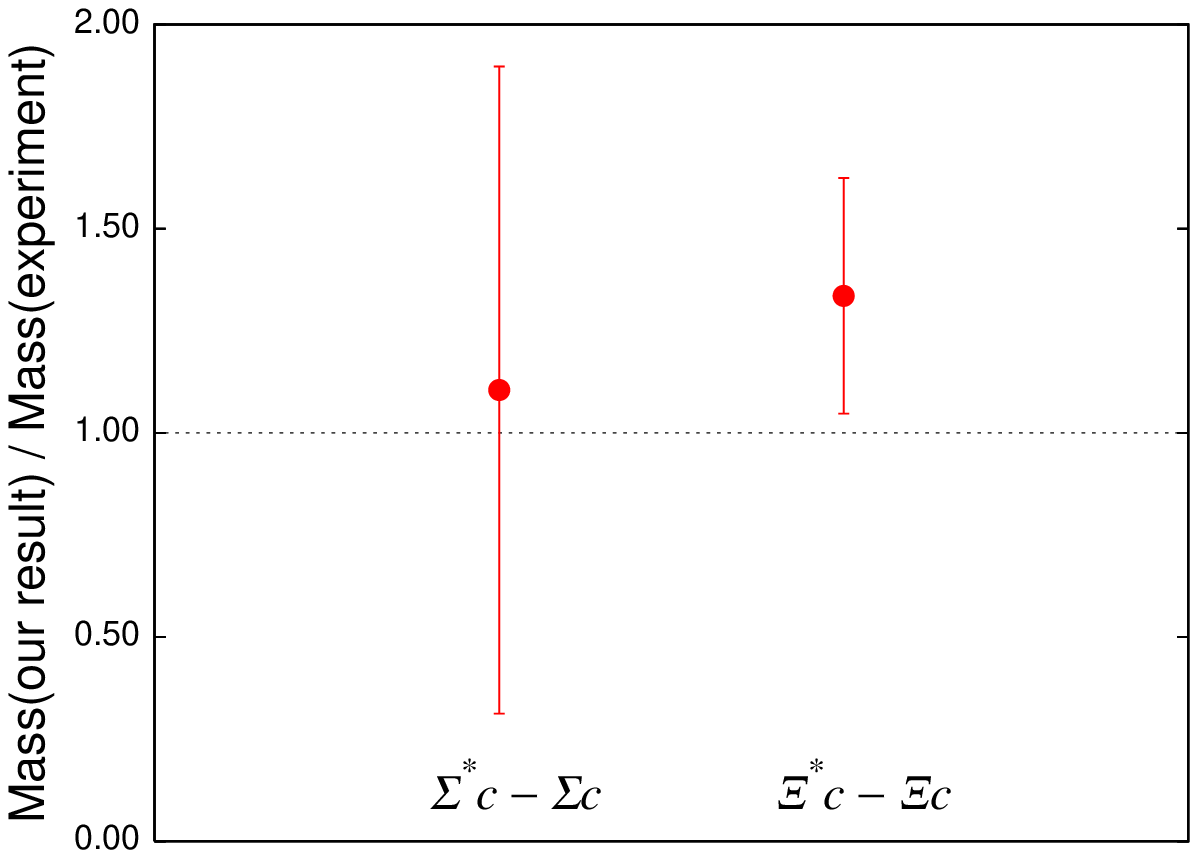}
 \caption{
  Comparison of mass differences of
  $\Sigma_c   - \Lambda_c$ types (left panel),
  $\Sigma_c^* -  \Sigma_c$ types (right panel).
 }
 \label{figure:mass_Sigma_c_Lambda_c_lattice}
\end{center}
\end{figure}

Figure~\ref{figure:mass_Sigma_c_Lambda_c_lattice}
displays mass differences.
Our results are consistent with experimental values within 2 $\sigma$ uncertainty.
These agreements indicate that 
the decomposition of $J=1/2$ $\Sigma$-type and $\Lambda$-type baryons,
as well as that of $J=3/2$ and $J=1/2$ charmed baryons,
have been made successfully in our calculation.

We note that several systematic errors
have not been fully evaluated yet for these results.
First, finite size effects are not taken into account.
Although the NLO heavy baryon chiral perturbation theory predicts that
finite size effects for charmed baryons are less than 1 \%,
higher order terms may give significant contributions.
A direct confirmation in lattice QCD
by comparing spectra among different lattice volumes is desirable.
Second, strong decays such as $\Sigma_c \rightarrow \Lambda_c \pi$
are not taken into account in our analysis,
since $\Sigma_c \rightarrow \Lambda_c \pi$ is kinematically prohibited
on our lattice volume.
Last but not least, our results are obtained at a single lattice spacing
without continuum extrapolation.
Although a naive order counting gives a percent level of cutoff effects from
$O( \alpha_s^2 f(m_Q a)(a \Lambda_{QCD}), f(m_Q a)(a \Lambda_{QCD})^2 )$ terms
in the relativistic heavy quark action,
the continuum extrapolation is necessary to remove this uncertainty.
Additional calculations should be performed in the future
to remove all systematic errors mentioned above.

\setcounter{equation}{0}
\section{Doubly and triply charmed baryon spectrum}
\label{section:doubly_triply_charmed_baryon}

For doubly and triply charmed baryons,
an experimental value has been reported only for $\Xi_{cc}$,
though the experimental status is controversial.
In the other channels, lattice QCD result gives predictions
before experimental mass measurements.

\begin{figure}[t]
\begin{center}
 \includegraphics[width=75mm]{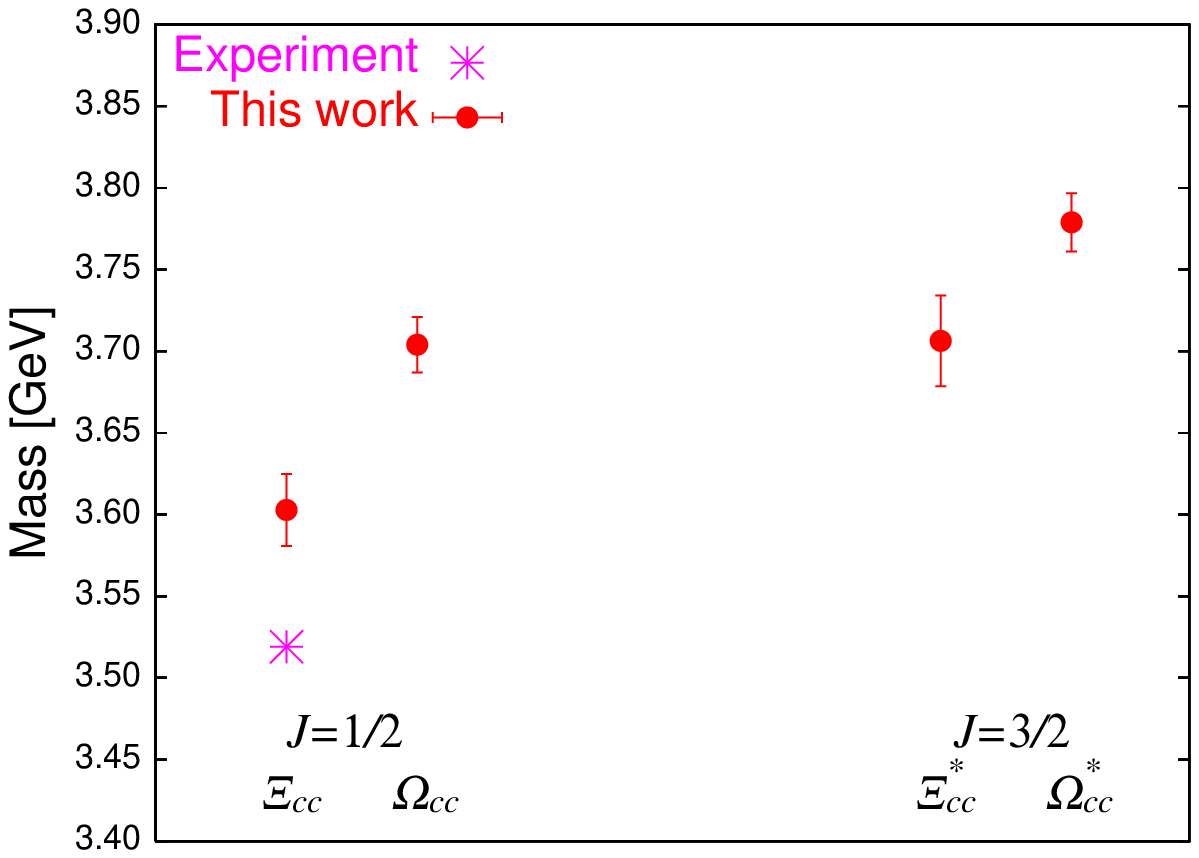}
 \includegraphics[width=75mm]{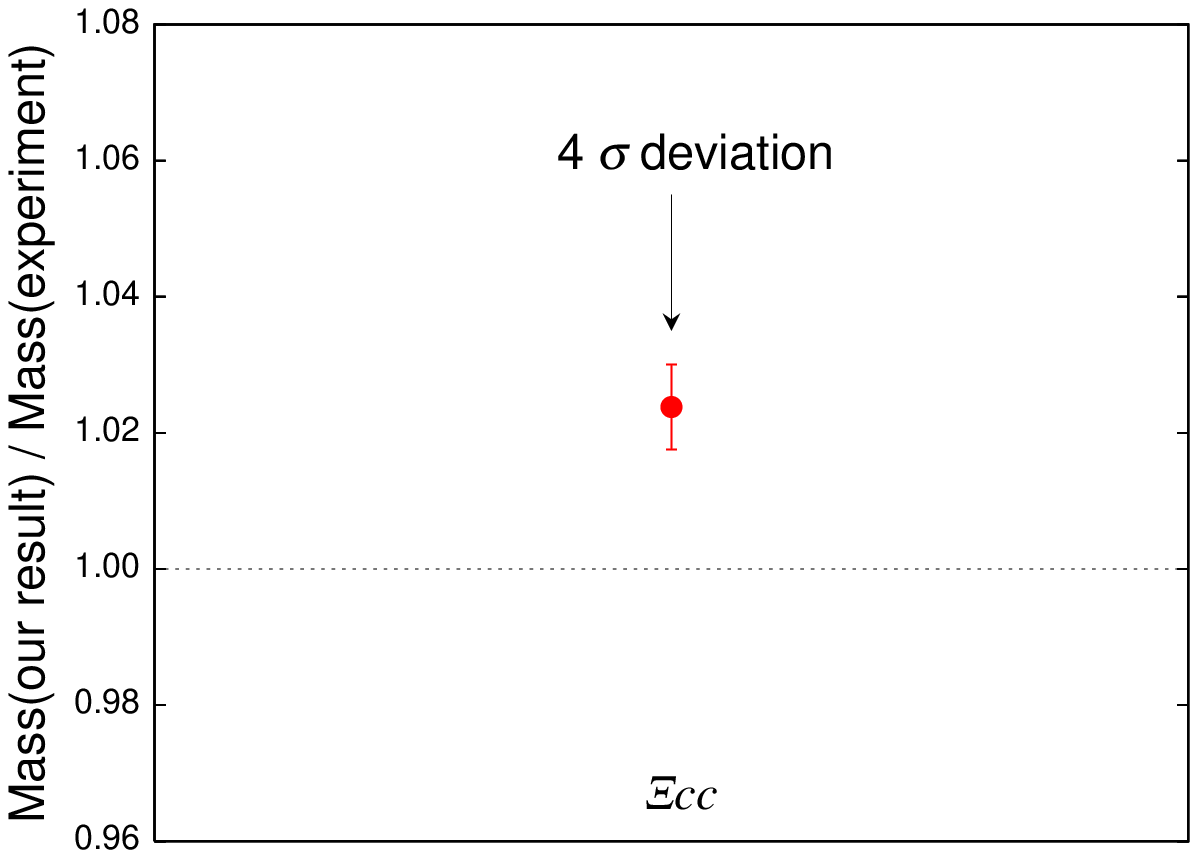}
 \caption{
 Our results for the doubly charmed baryon spectrum (left panel),
 and those normalized by the experimental value (right panel).
 }
 \label{figure:mass_doubly_charmed_experiment}
\end{center}
\end{figure}

\begin{figure}[t]
\begin{center}
 \includegraphics[width=75mm]{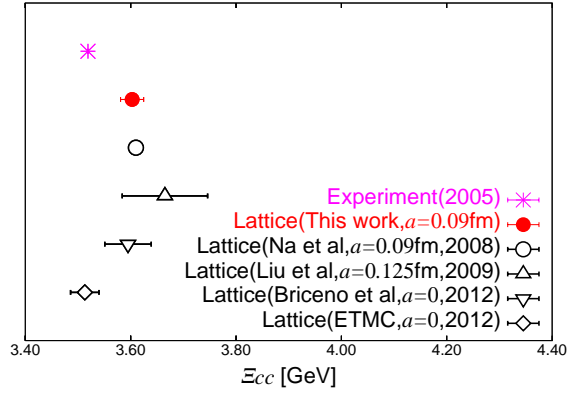}
 \caption{
 Comparison of $\Xi_{cc}$.
 }
 \label{figure:mass_Xi_cc_lattice}
\end{center}
\end{figure}

Figure~\ref{figure:mass_doubly_charmed_experiment} shows
our results for the doubly charmed baryons.
Our estimate for $m_{\Xi_{cc}}$ clearly deviates
from the experimental value
by SELEX Collaboration~\cite{SELEX_2002_2005}.
The difference is $4 \sigma$, as shown in the right figure.
Our result for $m_{\Xi_{cc}}$ is consistent with 
results from other lattice QCD calculations except ETMC,
as shown in Fig.~\ref{figure:mass_Xi_cc_lattice}.
This discrepancy is need to be understood and should be resolved.

\begin{figure}[t]
\begin{center}
 \includegraphics[width=75mm]{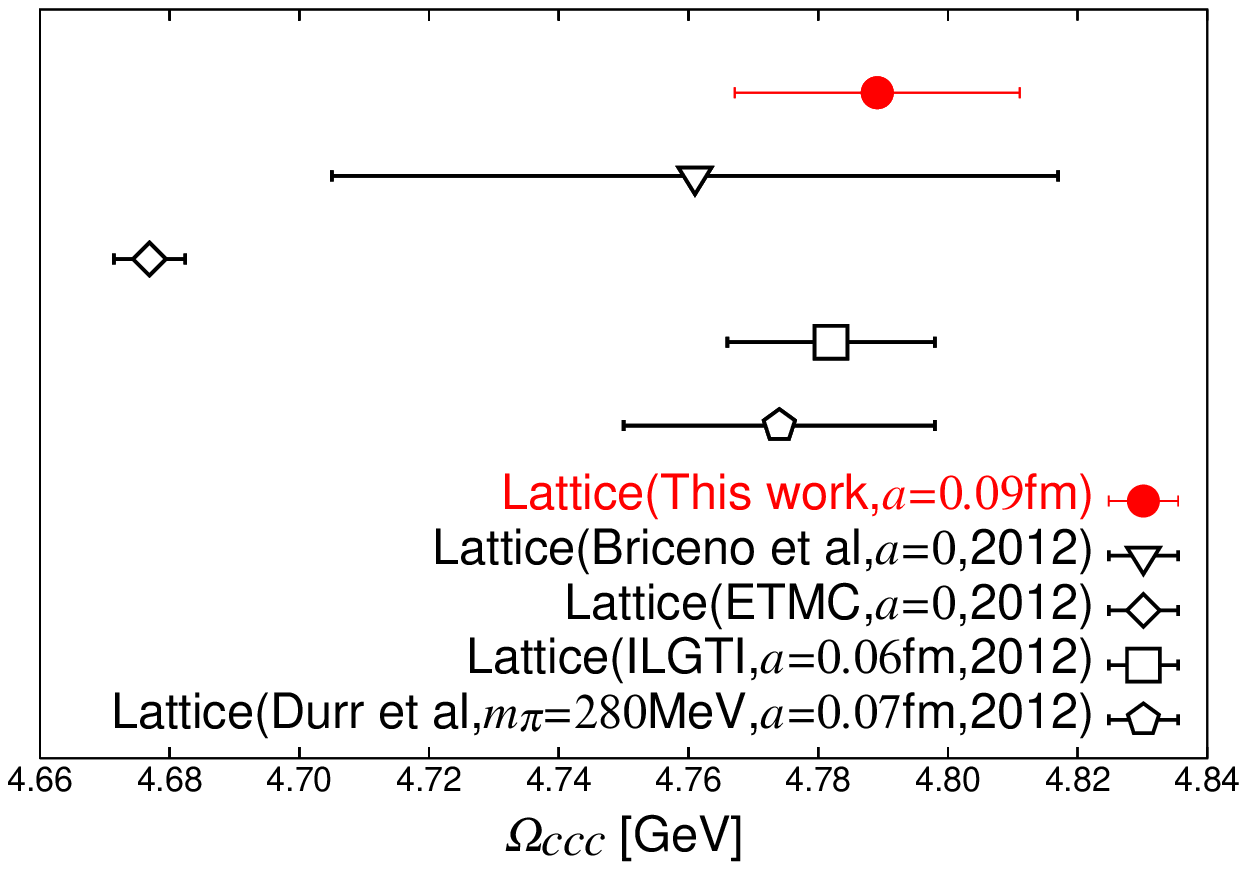}
 \includegraphics[width=75mm]{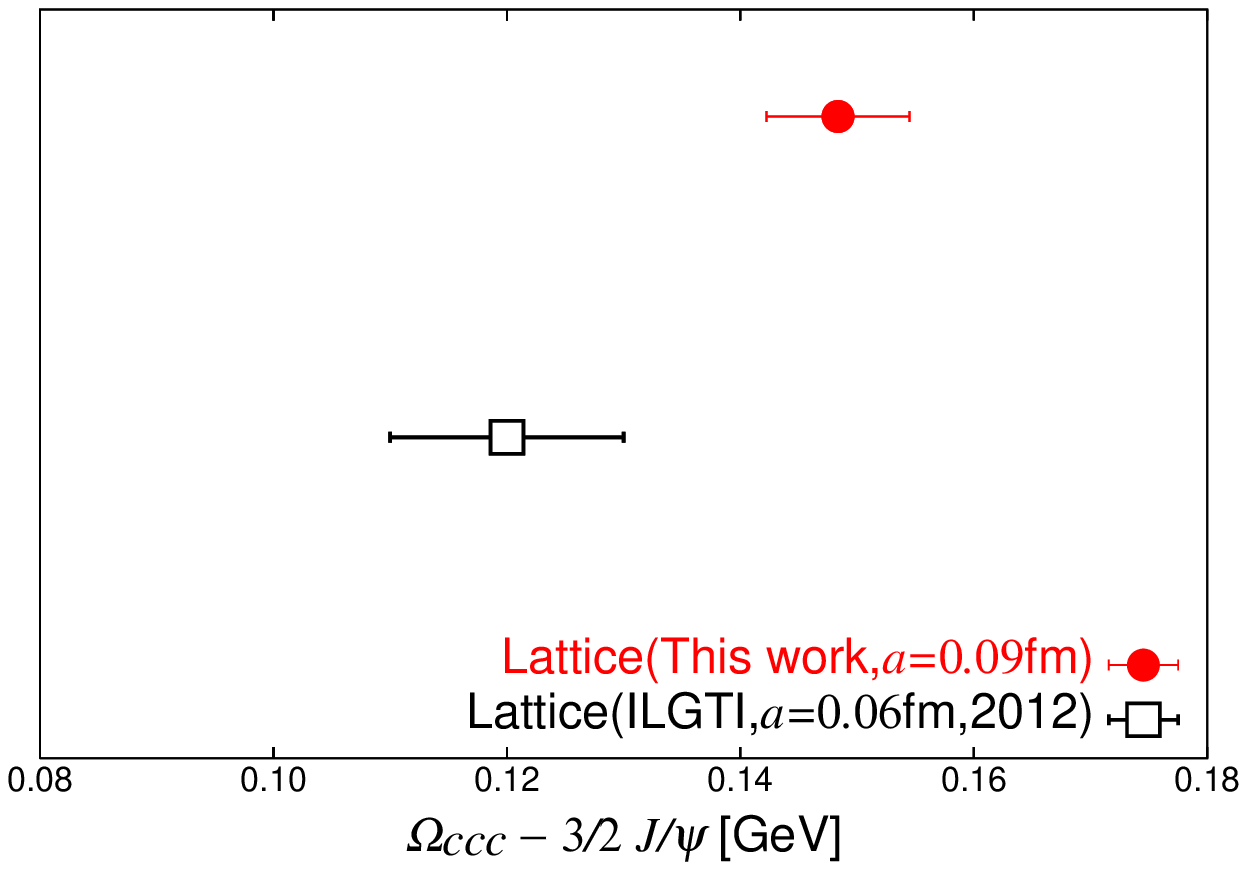}
 \caption{
 Comparison of $\Omega_{ccc}$ (left panel),
 and $\Omega_{ccc}$ $-$ $3/2$ $J / \psi$ (right panel).
 }
 \label{figure:mass_Omega_ccc_lattice}
\end{center}
\end{figure}

Similarly,
Fig.~\ref{figure:mass_Omega_ccc_lattice}
displays lattice QCD results for the triply charmed baryon
from several groups.
Our prediction agrees with that by others except ETMC.
A marginal discrepancy is observed between our value and that of Ref.~\cite{Mathur_2012}
for $m_{\Omega_{ccc}}$ $-$ $3/2$ $m_{J / \psi}$ mass difference (the right figure).

For a more detailed comparison,
precise evaluations of all systematic errors are required.
In particular, the largest source of systematic errors for our calculations
is the lattice artifact, which should be removed by the continuum extrapolation.

\setcounter{equation}{0}
\section{Conclusions}
\label{section:conclusions}

We have studied charmed baryon masses
in $N_f=2+1$ dynamical lattice QCD
at a lattice spacing of $a^{-1}=2.194(10)$~GeV.
The reweighting technique allows us
to perform a measurement
directly at the physical point.
It removes a systematic error associated with
chiral extrapolations of charmed baryon masses,
which had prevented previous lattice QCD calculations from 
predicting precise values for charmed baryon masses.

Our results for the mass spectrum of singly charmed baryons
are consistent with experiments within 2 $\sigma$ uncertainty.
This confirms that we are able to control charm quark mass corrections successfully,
not only in the meson sector~\cite{PACS_CS_2011}
but also in the baryon sector,
by use of the relativistic heavy quark action of Ref.~\cite{RHQ_action_Tsukuba}.

We then extract predictions for doubly charmed baryons.
Our result for $m_{\Xi_{cc}}$ is consistent
with values of other lattice QCD calculations 
employing the dynamical staggered quarks,
but disagree with the estimation by the ETMC.
Moreover, our $\Xi_{cc}$ mass
is different from the SELEX experimental value,
approximately by 85 MeV,
which corresponds to $4 \sigma$.
A similar deviation between ETMC and us is also observed
in the triply charmed baryon mass, $m_{\Omega_{ccc}}$.
Precise estimations of all systematic errors, especially the lattice artifacts,
are required to resolve these discrepancies.

\section*{Acknowledgments}

Y.N. thanks Yasumichi Aoki, Carleton DeTar and Nilmani Mathur
for valuable discussions,
and Stephan D\"{u}rr for his comment on this manuscript.
Numerical calculations for the present work have been carried out
on the PACS-CS computer
under the ``Interdisciplinary Computational Science Program'' of
the Center for Computational Sciences, University of Tsukuba.
This work is supported in part by Grants-in-Aid of the Ministry
of Education, Culture, Sports, Science and Technology(MEXT)-Japan
(Nos.~18104005,  20340047, 20540248, 21340049, 22244018, and 24540250),
the Grants-in-Aid for Scientific Research on Innovative Areas
(No. 2004: 20105001, 20105002, 20105003, and 20105005),
and SPIRE (Strategic Program for Innovative REsearch).


\end{document}